\documentclass[a4paper]{jpconf}
\newcommand{\beq}{\begin{eqnarray}}
\newcommand{\eeq}{\end{eqnarray}}

\newcommand{\be}{\mbox{$^9$Be}}
\newcommand{\he}{\mbox{$^6$He}}
\newcommand{\li}{\mbox{$^{11}$Li}}
\newcommand{\aan}{\mbox{$\alpha+\alpha+n$}}
\newcommand{\bepb}{\mbox{$^9$Be+$^{208}$Pb}}
\newcommand{\ben}{\mbox{$^8{\rm Be}+n$}}
\newcommand{\ahe}{\mbox{$^5{\rm He}+\alpha$}}

\usepackage{citesort}
\usepackage[dvips]{graphicx}
\usepackage{epsfig}
\usepackage{amsmath}

\begin{document}
\title{Four-body effects on $\bepb$ scattering and fusion around the Coulomb barrier
}

\author{P Descouvemont$^1$, T Druet$^1$, L F Canto$^{2,3}$ and M S Hussein$^{4,5,6}$}
\address{$ˆ1$ Physique Nucl\'eaire Th\'eorique et Physique Math\'ematique, C.P. 229,
Universit\'e Libre de Bruxelles (ULB), B 1050 Brussels, Belgium}
\address{$ˆ2$ Instituto de F\'isica, Universidade Federal do Rio de Janeiro, C.P. 68528, 21941-972 Rio de Janeiro, RJ, Brazil}
\address{$ˆ3$ Instituto de Física, Universidade Federal Fluminense, Av. Gal. Milton Tavares de Souza s/n, Niterói, RJ, Brazil }
\address{$ˆ4$ Departamento de F\'isica Matem\'atica, Instituto de F\'isica, Universidade de S\~ao Paulo, C.P. 66318, 05314-970, S\~ao Paulo, SP, Brazil}
\address{$ˆ5$ Instituto de Estudos Avan\c{c}ados, Universidade de S\~ao Paulo, C.P. 72012, 05508-970, 
S\~ao Paulo, SP, Brazil}
\address{$ˆ6$ Departamento de F\'{i}sica, Instituto Tecnol\'{o}gico de Aeron\'{a}utica, CTA, S\~{a}o Jos\'{e} dos Campos, S\~ao Paulo, SP, Brazil}
\ead{pdesc@ulb.ac.be}

\begin{abstract}
We investigate the $\bepb$ elastic scattering and fusion at energies around the Coulomb barrier. 
The $^9$Be nucleus is described in a $\aan$ three-body model, using the hyperspherical coordinate
method. The scattering with $^{208}$Pb is then studied with the
Continuum Discretized Coupled Channel (CDCC) method, where the $\aan$ continuum is approximated by a
discrete number of pseudostates. Optical potentials for the 
$\alpha+^{208}$Pb and $n+^{208}$Pb systems are taken from the literature. We present elastic-scattering
and fusion cross sections at different energies, and investigate the convergence with respect to
the truncation of the $\aan$ continuum. A good agreement with experiment is obtained, considering that there is no 
parameter fitting. We show that continuum effects increase at low energies. 
\end{abstract}

\section{Introduction}

Many experiments have been performed with the $\be$ nucleus, used as a 
target or as a projectile \cite{KAK09}.  Although $\be$ is stable, it presents a 
Borromean structure, as the well known halo nucleus $\he$.  None of the 
two-body subsystems $\alpha+n$ or $\alpha+\alpha$ is bound in $\be$,  
which has important consequences on the theoretical description of 
this nucleus.  Precise wave functions must include the three-body nature 
of $\be$.  The hyperspherical formalism \cite{ZDF93} is an ideal tool to 
describe three-body Borromean systems, as it does not assume a specific 
two-body structure, and considers the three particles $\alpha+\alpha+n$ on 
an equal footing.  

In the present work, we aim at investigating $\be$ scattering and fusion on an heavy
target.  The reaction framework is the Continuum Discretized Coupled 
Channel (CDCC) method \cite{Ra74b,YIK86,YMM12}, which is well adapted 
to weakly bound projectiles since it allows to include breakup channels.  
Over the last decades, the CDCC method has been extended in various directions, and in particular to reactions involving three-body projectiles such as $\he$ \cite{MHO04} or $\li$ \cite{CFR12}.  Going from two-body projectiles (such as d=p+n or $^7$Li=$\alpha$+t) to three-body 
projectiles strongly increases the complexity of the calculations, 
even if both options eventually end up with a standard 
coupled-channel system.  

Many data have been obtained for $\bepb$ elastic scattering 
\cite{WFC04,PJM11} and fusion 
\cite{DHB99,DGH04}.  These experimental data provide a good opportunity to test 
$\be$ wave functions.  Previous CDCC calculations, using a two-body approximation for $\be$, show that breakup
effects are important \cite{KKR01,PJM11,PJP13}. In a two-body model, $\be$ is assumed to have a
$\ben$ or $\ahe$ cluster structure. This approximation presents several shortcomings: $(i)$
$^8$Be and $^5$He are unbound, and assuming a point-like structure is questionable; $(ii)$ in a more
rigorous three-body approach, both configurations are strictly equivalent, and the relative importance
of the $\ben$ and $\ahe$ channels is not relevant; $(iii)$ a two-body model of $\be$ requires
$^8$Be+$^{208}$Pb and $^5$He+$^{208}$Pb optical potentials, which are not available. These different issues are more precisely addressed in a three-body model of $\be$. No assumption should be made about the cluster
structure, and $\alpha+^{208}$Pb as well as $n+^{208}$Pb optical 
potentials are available in the literature.

\section{Three-body model of $\be$}
\label{sec2}
The determination of the $\be$ wave functions is the first step for 
the $\bepb$ CDCC calculation.  The spectroscopy of $\be$ in cluster 
models has been performed in previous microscopic \cite{De01,ADB03,AOS96}
 and non-microscopic \cite{VKP95,TBD06b,AJG10} models.
For a three-body system, the Hamiltonian is given by
\beq
H_0=\sum_{i=1}^3 \frac{\pmb{p}^2_i}{2m_i}+\sum_{i<j=1}^3 V_{ij}
(\pmb{r}_i-\pmb{r}_j),
\label{eq1}
\eeq
where $\pmb{r}_i$ and $\pmb{p}_i$ are the space and momentum coordinates 
of the three particles with masses $m_i$, and $V_{ij}$ a potential between nuclei $i$ and $j$.  
For the $\alpha+\alpha$ interaction, we use the deep potential of 
Buck {\sl et al.} \cite{BFW77}.  The $\alpha$+n interaction is taken from 
Kanada {\sl et al.} \cite{KKN79}.  Both (real) potentials accurately reproduce 
the $\alpha+\alpha$ and $\alpha$+nucleon phase shifts over a wide 
energy range.  

With these potentials, the $\be$ ground state is too bound 
($-3.12$ MeV, while the experimental value is $-1.57$ MeV with respect 
to the $\aan$ threshold).  We have therefore introduced a phenomenological three-body force as
\beq
V_{3B}(\rho)=\frac{V_0}{1+(\rho/\rho_0)^2},
\label{eq9}
\eeq
where $\rho_0=6$ fm \cite{TDE00}. Using $V_0=2.7$ MeV provides the experimental 
binding energy of the $3/2^-$ ground state. 
The r.m.s. radius, the quadrupole moment, and the magnetic moment are $\sqrt{r^2}=2.36$ fm,
$Q(3/2^-)=4.96$ $e$.fm$^2$, and $\mu =-1.33\ \mu_N$, respectively. These values are in
fair agreement with experiment ($2.45\pm 0.01$ fm, $5.29\pm 0.04\ e$.fm$^2$ and $-1.18\ \mu_N$,
respectively) 

\section{The CDCC method}
\label{sec3}
We present here a brief outline of the CDCC method, and we refer to 
Ref.~\cite{RAG08} for specificities of three-body projectiles.  
The CDCC method is based on approximate solutions of the projectile Hamiltonian (\ref{eq1})
\beq
H_0 \Phi^{jm\pi}_k=E^{j\pi}_{0,k}\Phi^{jm\pi}_k,
\label{eq13}
\eeq
where $k$ are the excitation levels in partial wave $j\pi$.  
Solutions with $E^{j\pi}_{0,k}<0$ correspond to bound states of 
the projectile, whereas $E^{j\pi}_{0,k}>0$ correspond to narrow 
resonances or to approximations of the three-body continuum.  
These states cannot be associated with physical states, 
and are referred to as pseudostates.  

\begin{figure}[htb]
\begin{center}
\epsfig{file=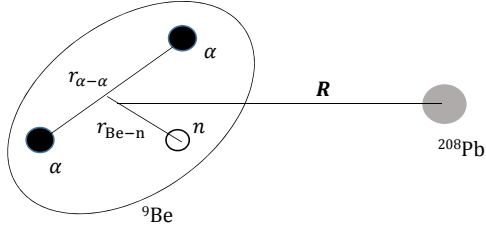,width=0.4\textwidth}
\caption{Coordinates involved in the present four-body model.}
\label{fig3}
\end{center}
\end{figure}

The Hamiltonian of the projectile + target system is written as
\beq
H(\pmb{R},\pmb{x},\pmb{y})=H_0(\pmb{x},\pmb{y})-\frac{\hbar^2}{2\mu_{PT}}
\Delta_R +V(\pmb{R},\pmb{x},\pmb{y})
\label{eq14}
\eeq
where $\mu_{PT}$ is the reduced mass of the system, and $\pmb{R}$ the relative coordinate
(see Fig.\ \ref{fig3}). The Jacobi coordinates $\pmb{x}$ and $\pmb{y}$ are proportional to 
$\pmb{r}_{\rm Be-n}$ and $\pmb{r}_{\alpha-\alpha}$, 
respectively. The potential term reads
\beq
V(\pmb{R},\pmb{x},\pmb{y})=V_{1t}(\pmb{R},\pmb{y})
+V_{2t}(\pmb{R},\pmb{x},\pmb{y})+V_{3t}(\pmb{R},\pmb{x},\pmb{y})
\label{eq15}
\eeq
where the three components $V_{it}$ are optical potentials between 
fragment $i$ and the target.  

In order to solve the Schr\"{o}dinger equation associated with 
(\ref{eq14}), the total wave function with angular momentum $J$ and parity $\Pi$ is expanded as
\beq
\Psi^{JM\Pi}_{(\omega)}(\pmb{R},\pmb{x},\pmb{y})=\sum_{j\pi kL}\,
\varphi^{JM\Pi}_{j\pi kL}(\Omega_R,\pmb{x},\pmb{y})
g^{J\Pi}_{jkL(\omega)}(R),
\label{eq16}
\eeq
where $\omega$ is the entrance channel, and where the channel wave functions are given by
\beq
\varphi^{JM\Pi}_{j\pi kL}(\Omega_R,\pmb{x},\pmb{y})=i^L
\biggl[\Phi^{j\pi}_k(\pmb{x},\pmb{y})\otimes
Y_L(\Omega_R)
\biggr]^{JM}.
\label{eq17}
\eeq
The radial functions $g^{J\Pi}_{c}(R)$ (we use the notation $c=(j,\pi,k,L)$) 
are obtained from the system
\beq
\biggl[ -\frac{\hbar^2}{2\mu_{PT}}\left( \frac{d^2}{dR^2}-\frac{L(L+1)}{R^2}\right) +E_{c}-E \biggr] 
g^{J\Pi}_{c(\omega)}(R) 
+\sum_{c'}V^{J\Pi}_{c,c'}(R) g^{J\Pi}_{c'(\omega)}(R)=0,
\label{eq18}
\eeq
where $E$ is the c.m. energy, and where the coupling potentials are given by matrix elements
\beq
V^{J\Pi}_{c,c'}(R)=\langle \varphi^{JM\pi}_{c} \vert V \vert 
\varphi^{JM\pi}_{c'} \rangle.
\label{eq19}
\eeq
As the fragment-target potentials (\ref{eq15}) are optical potentials, matrix elements (\ref{eq19}) contain
a real and an imaginary parts.
The integration is performed over $\Omega_R$ and over the internal coordinates $\pmb{x}$ and $\pmb{y}$.  The calculation of these 
coupling potentials is much more complicated for three-body projectiles 
than for two-body projectiles.  Out of the three potentials $V_{it}$ 
in (\ref{eq15}), matrix elements of $V_{1t}$ are the easiest since 
$V_{1t}$ does not depend on $\pmb{x}$.  The calculations of the 
two remaining terms $V_{2t}$ and $V_{3t}$ are performed by using the 
Raynal-Revai coefficients (see Refs. \cite{RAG08,TND04} for detail).
We solve Eq.\ (\ref{eq18}) with the $R$-matrix 
theory \cite{DB10}.

\section{Application to the $\bepb$ system}
\label{sec4}
\subsection{Elastic scattering}

\begin{figure}[htb]
\begin{center}
\epsfig{file=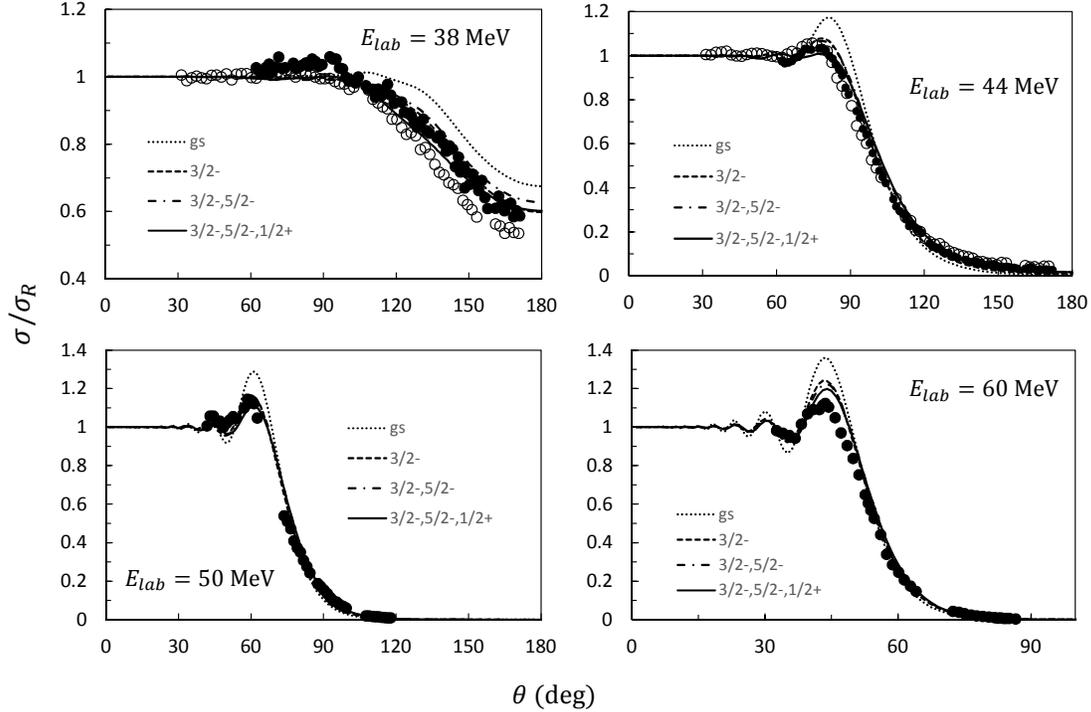,width=0.9\textwidth}
\caption{$\bepb$ elastic cross sections (divided by the Rutherford cross section)
 at different $\be$ laboratory energies, and for different sets of $\be$ partial waves. 
The experimental data are taken from
Ref.~\cite{WFC04} (filled circles) and Ref.~\cite{YZJ10} (open circles).}
\label{fig5}
\end{center}
\end{figure}

The calculations are performed with the $j^{\pi}=3/2^-,5/2^-,1/2^+$ partial waves on $\be$, and
the cutoff energy is 10 MeV.
Figure \ref{fig5} shows $\bepb$ elastic cross sections at various energies 
(the data are taken from Refs.\ \cite{WFC04,YZJ10}).  In contrast with 
optical-model calculations \cite{WFC04}, no renormalization of the potential 
is needed.  The absorption is simulated by the imaginary parts of the 
$\alpha-^{208}$Pb and n$-^{208}$Pb interactions, and by the $\alpha+\alpha+n$ 
discretized continuum.  From Fig. \ref{fig5}, it appears that continuum couplings 
are more important at low energies.  At $E=38$ and 44 MeV, the single-channel 
calculation, limited to the $\be$ ground state (dotted lines) is significantly 
different from the data.  This confirms the conclusion of a previous work \cite{DD12b} 
which suggests that continuum couplings are more important at low energies.

Calculations involving $3/2^-$ pseudostates are shown as dashed lines, 
and the full calculations as solid lines.  The effect of the $1/2^+$ partial wave 
is minor, and the main differences stem from the $5/2^-$ pseudostates.  

\subsection{Fusion}
The $\bepb$ fusion reaction has been studied theoretically by several groups
(see Ref.~\cite{JPK14} and references therein). Data about the various mechanisms (total TF,
complete CF and incomplete ICF fusion) are also available in the literature 
\cite{DHB99,DGH04,DHS10}.

The total fusion cross section $\sigma_{TF}$ is defined from
\cite{CGD06,HT12,CH13}
\beq
\sigma_{TF}(E)=\frac{\pi}{k^2}\sum_{J\Pi} (2J+1)T^{J\Pi}(E),
\eeq
where $k$ is the wave number, 
and $T^{J\Pi}$ is a transmission coefficient in partial wave $J\Pi$.
It can be obtained from the scattering matrices as
\beq
T^{J\Pi}(E)=\frac{1}{(2I_1+1)(2I_2+1)}\sum_{L_{\omega},L}
\bigl(\delta_{L_{\omega}L}-
\vert U^{J\Pi}_{\omega L_{\omega},\omega L}(E)\vert^2 \bigr),
\eeq
where $I_1$ and $I_2$ are the spins of the colliding nuclei. Using flux conservation properties, the transmission coefficient can be expressed from the
imaginary part $W^{J\Pi}_{c,c'}(R)$ of the coupling potentials \cite{CH13}. It is therefore strictly equivalent to
\beq
T^{J\Pi}(E)=-\frac{4k}{E}\frac{1}{(2I_1+1)(2I_2+1)}\sum_{L_{\omega}}\sum_{\alpha,\alpha',L,L'}
\int g^{J\Pi}_{\omega L_{\omega},\alpha L}(R)\,
W^{J\Pi}_{\alpha L,\alpha' L'}(R)\,
g^{J\Pi}_{\omega L_{\omega},\alpha' L'}(R) \, dR,
\label{eq_t}
\eeq
where index $\alpha$ stands for $\alpha=(j,\pi,k)$, i.e. it does not depend on the angular
momentum $L$ (in other words $c=(\alpha,L)$).
This derivation is more complicated since it requires the wave functions, but it provides a better
accuracy at low energies, where the scattering matrix is close to unity. It also provides
an approximate method to distinguish between CF and ICF  \cite{DT02}.

\begin{figure}[h]
\begin{minipage}[t]{0.45\textwidth}
\includegraphics[width=\textwidth]{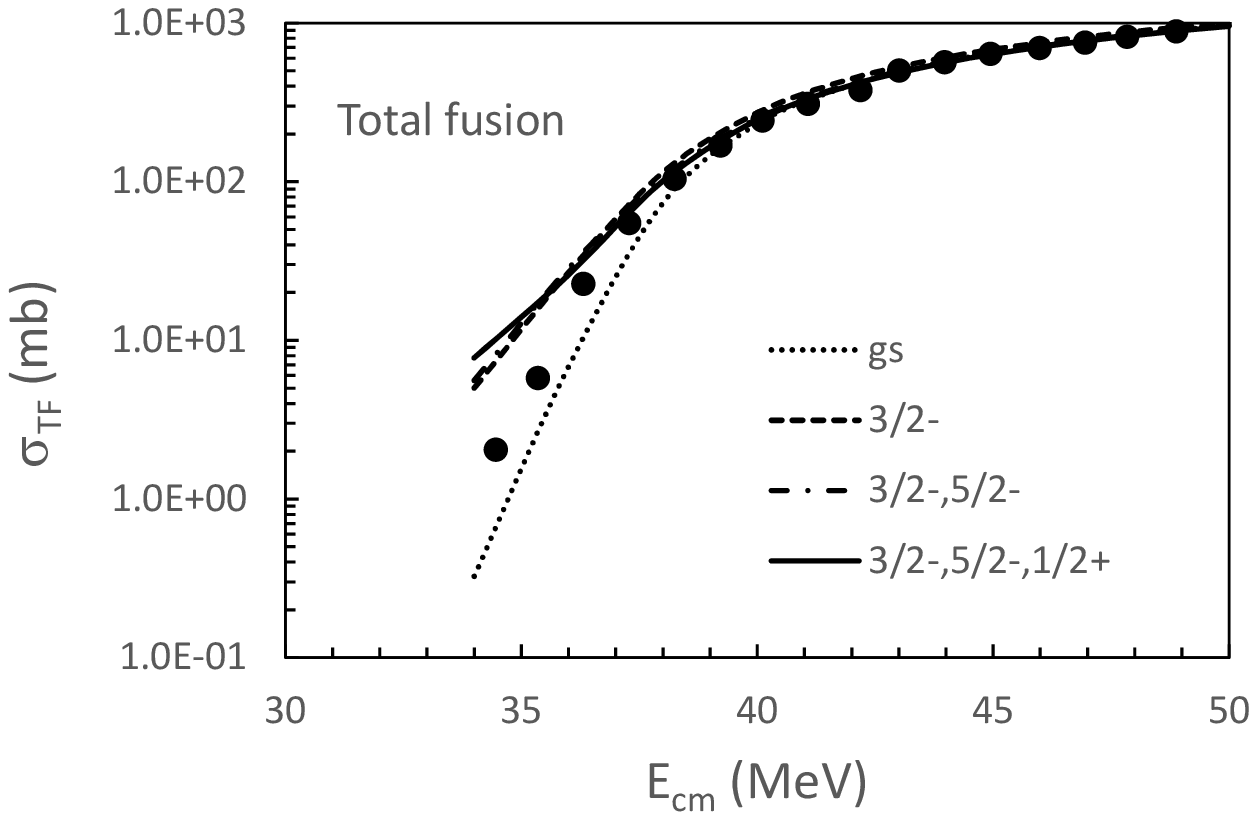}
\caption{\label{fig8}$\bepb$ total fusion cross section in a logarithmic scale. The data are taken from \cite{DHB99}.}
\end{minipage}
\hspace{0.1\textwidth}
\begin{minipage}[t]{0.45\textwidth}
\includegraphics[width=\textwidth]{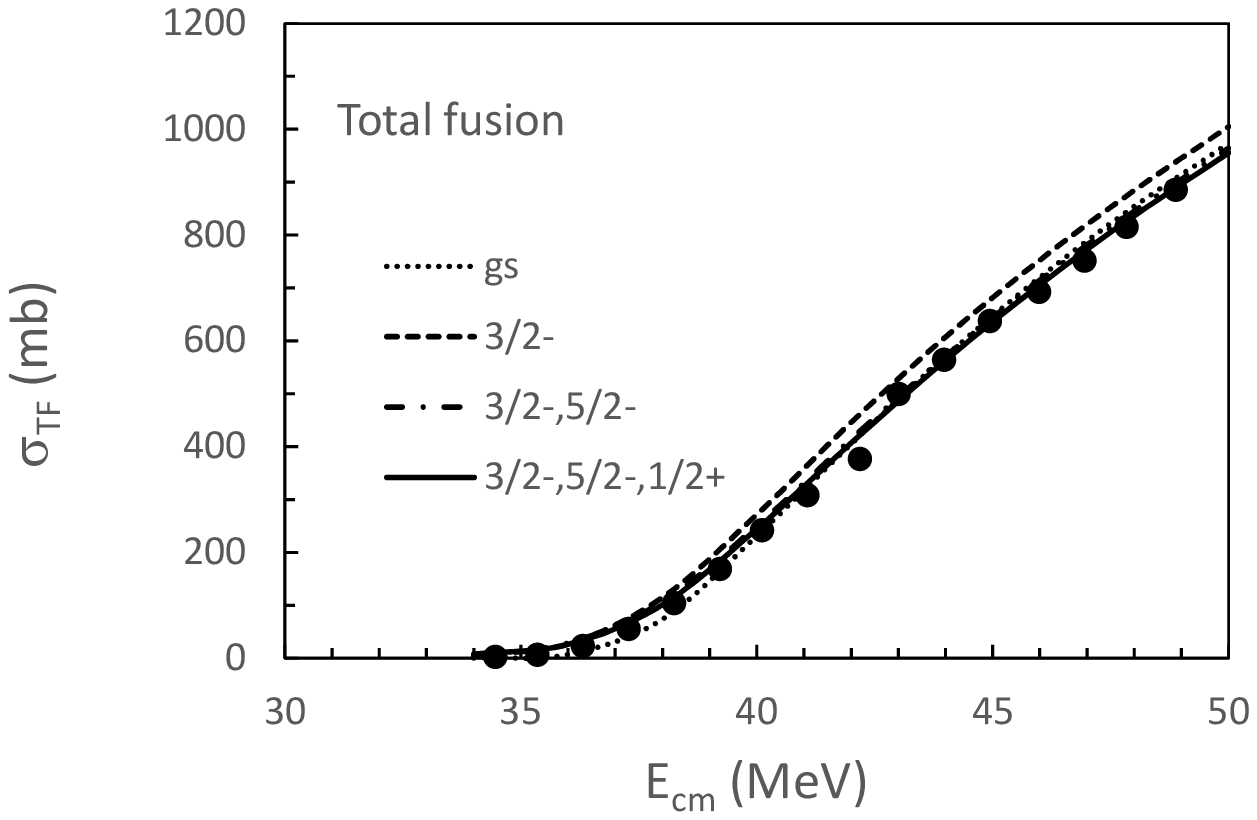}
\caption{\label{fig8b}See Fig.\ \ref{fig8} in a linear scale.}
\end{minipage} 
\end{figure}

The TF cross section is shown in  logarithmic (Fig.~\ref{fig8}) and linear (Fig.~\ref{fig8b}) scales. Convergence against
$\be$ states is rather fast, except below the Coulomb barrier where neglecting breakup channels leads to an underestimation of the cross section. Breakup effects, however, introduce a strong enhancement of the
cross section at low energies. Similar conclusions have been
drawn by Jha {\sl et al.} \cite{JPK14}.

\section{Conclusion}
\label{sec5}
We have applied the CDCC formalism to the $\bepb$ system. In a first step, we have computed 
$\be$ wave functions in a three-body $\aan$ model. These calculations were performed for low-lying states, but also for pseudostates, which represent positive-energy approximations
of the continuum. The main advantage of the hyperspherical approach is that it treats the
three-body continuum without any approximation concerning possible $\ben$ or $\ahe$ cluster
structures, which are in fact strictly equivalent.

The three-body model of $\be$ relies on $\alpha+\alpha$ and $\alpha+n$ (real) interactions,
which reproduce very well the elastic phase shifts. With these bare interactions, the $\be$ ground state is slightly too bound. We therefore introduce a phenomenological three-body
force to reproduce the experimental ground-state energy. The spectroscopic properties of low-lying
states is in fair agreement with experiment. 

We used first the $\be$ wave functions in a CDCC calculation of the $\bepb$ elastic scattering
at energies close to the Coulomb barrier.  As expected, including continuum channels
improves the theoretical cross sections. A fair agreement with the data is obtained.

We also applied the four-body  CDCC to the $\bepb$ fusion, by using the same potentials
as for elastic scattering. The total fusion cross section is found in good agreement with
experiment, except at very low energies, where we overestimate the data.

\section*{References}
%\bibliographystyle{iopart-num}
%\bibliography{biblio}
\providecommand{\newblock}{}

\end{document}